\documentclass[preprint,preprintnumbers,nofootinbib,amsmath,amssymb]{revtex4-1}
\usepackage{color}

\usepackage[active]{srcltx}

\usepackage{amsfonts}    % list packages between braces
\usepackage{amssymb}
\usepackage{xspace}
\usepackage{latexsym}
\usepackage{graphicx}

\usepackage[english]{babel}

\newcommand{\hh}{$\,{\rm H}_2\,\,$}

\newcommand{\al}{\alpha}

\newcommand{\Si}{\Sigma}

\newcommand{\si}{\sigma}

\begin{document}

\title{The molecule $\rm{H}_{2}$ in a strong magnetic field revisited}

\author{D. J.~Nader}
\email{djulian@uv.mx}
\affiliation{Facultad de F\'isica, Universidad Veracruzana,\\ Apartado Postal 91-090,
Xalapa, Veracruz, M\'exico}

\author{A.V.~Turbiner}
\email{turbiner@nucleares.unam.mx}
\affiliation{Instituto de Ciencias Nucleares, Universidad Nacional Aut\'onoma de M\'exico,\\ Apartado Postal 70-543,
04510 Ciudad de M\'exico, M\'exico}

\author{J.C.~L\'opez Vieyra}
\email{vieyra@nucleares.unam.mx}
\affiliation{Instituto de Ciencias Nucleares, Universidad Nacional Aut\'onoma de M\'exico,\\ Apartado Postal 70-543,
04510 Ciudad de M\'exico, M\'exico}

\begin{abstract}
A compact, few-parametric, physically adequate, 3-term variational trial function is used to calculate with high accuracy the energy of the ground state ${}^3\Pi_u$ of the hydrogen molecule \hh in strong magnetic field ${\bf B}$ in the range \mbox{$5\times10^{10}\, {\rm G} \leq B \leq 10^{13}\,$G.} The nuclei (protons) are assumed as infinitely massive (BO appproximation of zero order) and situated along the magnetic field line (parallel configuration).
\end{abstract}

\maketitle

\section{Introduction}

It is well known that as a consequence of the compression of the magnetic field flux during a supernova explosion, the extremely compact remnant object called neutron star  ~\cite{1967Natur.216..567P,Bignami:2003} can be formed having enormous magnetic fields on its surface. Typically the surface magnetic fields observed for neutron stars are in range  \hbox{$B=10^{10}-10^{12}\,$G} but can reach $\sim 10^{16}\,$G for magnetars.   Some neutron stars are characterized by the atmosphere - a few-centimeters-thick transition domain from nuclear densities to vacuum. Content of the atmosphere is one of the interesting open questions. It can be made from electrons, protons, $\al$-particles and even heavier nucleus, which can be condensed into atomic-molecular type systems in spite of high surface temperature $\sim 10^{4-5}$\,K. The simplest scenario is when the atmosphere is made from electrons and protons thus leading as the result of condensation to a formation of hydrogenic species.

It is well known that under a strong magnetic field, the structure of atoms and molecules is essentially modified: they loose their traditional, field-free forms becoming more bound and more compact with magnetic field increases, the electronic cloud gets a pronounced cigar-like form, and eventually the nucleus are aligned on the magnetic line, leading possibly to formation of finite chains, with all electron spins antiparallel to the magnetic field direction (see \cite{Ruderman:1971}, \cite{Kadomtsev:1971a} for a qualitative description and  \cite{Turbiner:2006}, \cite{Nader:2019}, references therein, for quantitative calculations).
The striking property of matter in strong magnetic fields is the existence of exotic systems for  sufficiently large magnetic field. A complete description of exotic atomic and molecular systems of one-two electrons which appear for magnetic fields, $B \lesssim B_{\rm Schw} = m^2 c^2/ e\hbar = 4.414 \times 10^{13}\, \rm G$, where the non-relativistic consideration holds, can be found at \cite{Turbiner:2006} and \cite{Turbiner:2010}, respectively, see also \cite{assci:2007}.

Among hydrogenic species (H-atom, H${}_2^+$,  H${}_3^+$,  H${}^-$ etc) the hydrogen molecule
${\rm H}_2$ plays prominent role due to its numerous appearances on the Earth: it is the simplest stable diatomic molecule. In weak magnetic fields $B \lesssim 0.1\,$a.u. ($1\,$a.u. $\equiv 2.35\times10^9\,$G) the ground state remains the spin-singlet state $^1\Si_g$ in the parallel configuration. Nonetheless, the ground state evolves as a function of the magnetic field strength in such a way that for magnetic fields in the range $0.18 \lesssim B \lesssim 12.3$\,a.u. the minimal energy state is realized by the (unbound) repulsive state $^3\Si_u$, see \cite{Detmer:1998}, it is represented by two hydrogen atoms in its ground states situated inifinitely far away from each other with spins antiparallel to magnetic field. Eventually, for magnetic fields larger than $B \gtrsim 12.3$\,a.u. ($\sim 3 \times 10^{10}$ G) the molecule
${\rm H}_2$ reappears as a stable system in the spin-triplet state ${}^3\Pi_u$.

It is worth mentioning that in the regime of strong magnetic fields $B \gtrsim 10^{11}\,$G, the hydrogen molecule ${\rm H}_2$ has been explored sparsely. Following the recent paper about 3-electron finite hydrogenic chains ${\rm H}_3$, ${\rm H}_2^-$ \cite{Nader:2019}, it is clear  that accurate study of the hydrogen molecule ${\rm H}_2$ in domain $B \sim 2 \times 10^{11}\,$G plays important role to understand their stability. Aim of this paper is study the hydrogen molecule ${\rm H}_2$ in the state $^3\Pi_u$ in a strong magnetic field in framework of the same variational method with compact trial function in the form of linear superposition of 3 terms. This trial function will allow us to check convergence with number of terms.

\section{The Hamiltonian and generalities}

The Hamiltonian of the hydrogen molecule ${\rm H}_2$ subject to uniform magnetic field ${\bf B}$   parallel to the molecular axis (coinciding with $z$-axis), which connect two infinitely massive  nuclei, is given by
\begin{widetext}
\begin{equation}
 \label{Ham}
 \mathcal{H}=-\sum_{i=1}^{2}\left(\frac{1}{2}\nabla^2_{i}+\sum_{\eta=A,B}\frac{1}{r_{i\eta}}\right)+
 \frac{1}{r_{12}}+\frac{B^2}{8}\sum_{i}^{2}\rho_{i}^2+\frac{B}{2}(L_{z}+2S_{z})
 +\frac{1}{R}\ ,
\end{equation}
\end{widetext}
when written in atomic units $\hbar=e=m_e=c=1$, where $\nabla_{i}$ with $i=1,2$, represents the 3-dimensional vector of the momentum of the $i$-th electron. Terms $-\frac{1}{r_{i\eta}}$ and $\frac{1}{r_{12}}$ are the Coulombic interactions between electron-nucleus (nuclei denoted as $\eta=A,B$) and electron-electron, respectively, where $r_{i\eta}$ is the distance between $i$-th electron and nucleus $\eta$, and $r_{12}$ is the distance between electrons.
In turn, the term $\frac{1}{R}$ is the classical repulsion between the (fixed) nuclei. In Hamiltonian (\ref{Ham}) the symmetric gauge (${\bf A}=\frac{1}{2}{\bf B}\times{\bf r}$) was chosen and therefore it includes the standard linear Zeeman term $\frac{1}{2}{\bf B}\cdot{\bf L}$ and diamagnetic  terms $\frac{B^2}{8}\rho_{i}^2\,$, where  $\rho_{i}^2=x_{i}^2+y_{i}^2$. The spin-Zeeman term  $\frac{g}{2}{\bf B}\cdot{\bf S}$, is also included (with   g-factor, $g=2$).

The conserved quantities of the system are following:
i) the projection $L_z$ of the total electronic angular momentum on the internuclear axis, giving rise to the magnetic quantum number  $M$,
ii) $S^2$ of the total electronic spin ($S=0,1$).
iii) the projection $S_z$ of the total electronic spin on the internuclear axis
iv) the $z$-parity \hbox{($\si=\pm 1$)} with respect to $z \to -z$.
These conserved quantities characterize the state of the molecule leading to the standard spectroscopic notation $^{2S+1}(M)_\pi$, where $\pi=\pm 1$ is the parity of the molecule $\pi=\sigma(-1)^M$, which is denoted as $g/u$ for positive/negative parity. The magnetic quantum number\\ $M=-1$ corresponds to $\Pi$ state.

\section{The method}

Following the physical relevance arguments described in \cite{Turbiner:1984,Turbiner:2006} the basic trial function is designed as a product of Landau orbitals, Slater orbitals for each electron-nucleus pair and correlation factor in exponential form,
\begin{equation}
\label{trial}
 \psi({\bf r}_1,{\bf r}_2;\{\al\}, \{\beta\})\ =\ \prod_{k=1}^2 (\rho_k^{|m_k|}e^{im_k\phi_k}e^{ -\alpha_{kA}r_{kA}-\alpha_{kB}r_{kB}-\frac{B\beta_k}{4}
 \rho_k^2})\,e^{\al_{12}r_{12}}
\end{equation}
where $\al_{kA,kB}$, $\beta_{k}$ with $k=1,2$ as well as $\al_{12}$ are non-linear variational parameters which have a physical meaning of screening (or anti-screening) factors (effective charges). The variables $\phi_k, k=1,2$ in (\ref{trial}) are the azimuthal angles of the position vector of each electron around the magnetic field line and $m_k$ are the individual magnetic quantum numbers ($m_1+m_2=M$). Thus, we use a trial function which is a superposition of three terms of the type (\ref{trial}):
\begin{equation}
 \label{superposition}
 \Psi=A_1\psi_1 + A_2 \psi_2+A_3\psi_3\ ,\quad \psi_i=\psi({\bf r}_1,{\bf r}_2;\{\al^{(i)}\}, \{\beta^{(i)}\})\ ,
\end{equation}
where $A_{1,2,3}$ are linear variational parameters, and each function $\psi_{i}$ has its own set of non-linear variational parameters $\al^{(i)}_{kA,kB}$, $\beta^{(i)}_{k}$,  $\al^{(i)}_{12}$.  In principle, each $\psi_{i}$ can describe a certain physical situation (see \cite{et2578,ALIJAH:2019} for details).
Without loss of generality $A_1=1$, therefore the total number of variational parameters is $24$
including the internuclear distance $R$ which is treated as an additional variational parameter.
Since the state of interest is $^3\Pi_u$ - it is of total spin $S=1$ and parity $\si=1$, the following symmetrizer
\begin{equation}
 \label{sym}
 (1+\hat{P}_{AB})(1-\hat{P}_{12})\ ,
\end{equation}
must be applied to each configuration (\ref{trial}) in (\ref{superposition}). Here, $\hat{P}_{AB}$ stands for the operator of permutation of nuclei, and $\hat{P}_{12}$ is the operator of permutation of the electrons.

%The idea behind Ansatz (\ref{superposition}) is to incorporate the different  mechanisms of %bonding one at the time into the trial function.
As a first step, a single function $\psi_{1}$ (\ref{trial}) is used as a trial function. The process of minimization then leads to a configuration of parameters corresponding to the dominant bonding mechanism. In the next step, a second function $\psi_{2}$  (\ref{trial}) is added to the first function $\psi_{1}$. This time, by keeping parameters of $\psi_{1}$ fixed the process of minimization determines the subdominant bonding mechanism defining $\psi_{2}$. Then the parameters
of $\psi_{2}$ are fixed and the parameters of $\psi_{1}$ readjusted. This process of fixing and releasing is repeated several times until convergence is reached. A similar procedure is performed when the third function $\psi_{3}$ is added to the linear superposition of $\psi_{1,2}$ (\ref{superposition}). The trial function  (\ref{superposition}) is essentially the generalization of the linear superposition used in \cite{et2578} to study the ground state  $^1\Si_g$ of the molecule ${\rm H}_2$ as well as for H${}_3^+$ \cite{TL:2013} in the field-free case and \cite{TurHe:2017}, \cite{ALIJAH:2019} in the presence of a magnetic field. It is important to note that after making minimization the variational parameters {\it vs} magnetic field strength are smooth slow-changing functions.

\section{Results}

The energies and the equilibrium distances of ${\rm H}_2$ in the $^3\Pi_u$ state obtained using the superposition of 3 configurations (\ref{superposition}) are presented in Table \ref{H2energies} for magnetic fields in the range $5 \times 10^{10}\leq B \leq 10^{13}\,$G.
For illustration of convergence with number of configurations,
we present the variational results obtained with 1, 2, 3 configurations. The results available in the literature \cite{Detmer:1998,Lai-Salpeter:1992,Lai-Salpeter:1996} are included for comparison.

Qualitatively, one can observe that the molecule ${\rm H}_2$ with a magnetic field growth becomes more bound (the binding energy increases) and more compact (the internuclear equilibrium distance decreases). For all values of the magnetic field the energy reduces when one more configuration is added to the superposition (\ref{superposition}) demonstrating convergence. Calculations show that by adding the second configuration to the first one reduces energy to $0.6-0.8 \%$, while adding the third configuration reduces energy by $0.1-0.2\%$. It indicates that the energy converges rapidly with number of configurations in the linear superposition (\ref{superposition}), similar rate of convergence was observed in \cite{Turbiner:2006,et2578,Nader:2019}).

%%%%%%%%%%%%%%%%%%%%
%%%%%%%%%%%%%%%%%%%%

Our variational energies for magnetic fields 50 and 100\,a.u differ from the most accurate ones available at present \cite{Detmer:1998} in $\sim 0.7\%$ obtained using very large basis of $1700-3300$ non-orthogonal optimized non-spherical Gaussian atomic orbitals. Note that the basis set of 207 terms the smaller than one used in \cite{Detmer:1998} was previously established in \cite{Detmer:1997} to study the singlet ground-state $^1\Si_g$ of H$_2$ molecule in field-free case. Notably, such a basis turns out to yield much less accurate energies than those obtained using a compact trial function for the state $^1\Si_g$ similar to (\ref{superposition}), see \cite{et2578}, and similarly for weak magnetic fields, see \cite{ALIJAH:2019}.
On another side, our total variational energies give comparable or lower values than those obtained by a multiconfigurational Hartree-Fock method \cite{Lai-Salpeter:1992}
for magnetic fields $B <2 \times 10^{12}\,$G.  However, for stronger magnetic fields the energies obtained by the Hartree-Fock method appear to be slightly lower than our variational energies, see Table~\ref{H2energies}, the difference occurs in the 4th figure. However, it is worth mentioning that in the later study \cite{Lai-Salpeter:1996}, but using a multiconfigurational Hartree-Fock method, only three figures of their previous energies were confirmed.
% Comparing the results of the present work with those in \cite{Lai-Salpeter:1996} they coincide, after rounding,  in all digits.
It is probable that the last digit in the energies presented in \cite{Lai-Salpeter:1992} was overestimated by the Hartree-Fock method.

%In any case, further independent studies are required to verify this guess.

%As a last remark, we observe that, in general, the variational parameters appearing  in (\ref{trial}-\ref{superposition}) have a smooth behavior as a function of the magnetic field and can be easily interpolated by means of low degree polynomials in $B$.

\widetext

 \begin{table}[h!]
\begin{center}
{\setlength{\tabcolsep}{0.2cm} \renewcommand{\arraystretch}{1}
 \resizebox{!}{0.5\textwidth}{%
         \begin{tabular}{c|cc|cc|cc}
\hline \hline
    & \multicolumn{2}{c|}{1 conf} & \multicolumn{2}{c|}{2 confs} & \multicolumn{2}{c}{3 confs} \\
     $B\,$($10^9\,$G) & $E$(a.u.) & $R_{eq}$(a.u.) & $E$(a.u.)   & $R_{eq}$(a.u.) & $E$(a.u.) & $R_{eq}$(a.u.) \\
\hline
%$47.0$           &            &        &           &          & $-4.493$   & $0.664$ \\
%            &            &        &           &          & $-4.519$   & $0.65$ \\
% $50.0$           &            &        &           &          & $-4.598$   & $0.653$ \\
$0$       &      &      &      &      & $-1.1742\,{}^{\dagger\ 1}$ & $1.40\,{}^{\dagger\ 1}$ \\
          &      &      &      &      & $-1.1734\,{}^{\dagger\ 2}$ & $1.40\,{}^{\dagger\ 2}$ \\
 $50.0$          &          &        &           &          & $-4.574$  & $0.720$ \\
 $100.0$         &          &        &           &          & $-5.972$  & $0.522$ \\
                 &          &        &           &          & $-5.89\,{}^a$ & $0.54\,{}^a$ \\
                 &          &        &           &          & $-5.92\,{}^b$ & $0.52\,{}^b$ \\
 $117.5$         & $-6.299$ & $0.480$ & $-6.336$ & $0.476$  & $-6.342$  & $0.476$ \\
                 &          &        &           &          & $-6.384\,{}^c$ & $0.48\,{}^c$ \\
 $235.0$         & $-8.113$ & $0.38$ & $-8.173$  & $0.385$  & $-8.178$  & $0.384$\\
                 &          &        &           &          & $-8.236\,{}^c$ & $0.38\,{}^c$ \\
% $282.0$         & $-8.6610$ & $0.363$& $-8.7281$ & $0.361$ & & \\
% $352.5$         & $-9.3763$ & $0.343$& $-9.4512$ & $0.335$ & & \\
 $500.0$         & $-10.598$ & $0.295$ & $-10.686$ & $0.296$ & $-10.695$  & $0.301$ \\
                 &           &         &           &         & $-10.67\,{}^a$ & $0.31\,{}^a$ \\
                 &           &         &           &         & $-10.7\,{}^b$  & $0.30\,{}^b$ \\
 $1000.0$        & $-13.443$ & $0.247$ & $-13.556$ & $0.245$ & $-13.576$  & $0.243$ \\
                 &           &         &           &         & $-13.55\,{}^a$ & $0.24\,{}^a$ \\
                 &           &         &           &         & $-13.6\,{}^b$  & $0.25\,{}^b$ \\
 % $2000.0$       & $-16.910$  & $0.202$& $-17.055$ & $0.200$  &            &          \\
 %                &            &        &           &          & $-17.10$   & $0.19$  \\
 %                &            &        &           &          & $-17.1$    & $0.20$  \\
 $2000$          &           &         &           &         & $-17.09$     & $0.197$ \\
                 &           &         &           &         & $-17.10\,{}^a$ & $0.19\,{}^a$ \\
                 &           &         &           &         & $-17.1\,{}^b$  & $0.19\,{}^b$ \\
 $2350.0$        & $-17.817$ & $0.191$ & $-17.969$ & $0.192$ & $-18.006$ & $0.188$ \\
                 & $-17.816\,{}^{d}$ & $0.190\,{}^{d}$       &  &  &   &  \\
 $5000$          &           &         &           &         & $-22.85$ & $0.151$ \\
                 &           &         &           &         & $-22.92\,{}^a$ & $0.15\,{}^a$ \\
                 &           &         &           &         & $-22.9\,{}^b$ & $0.15\,{}^b$ \\
 $10000$         &           &         &           &         & $-28.18$ & $0.126$ \\
                 &           &         &           &         & $-28.22\,{}^a$ & $0.12\,{}^a$ \\
                 &           &         &           &         & $-28.2\,{}^b$  & $0.12\,{}^b$ \\
 \hline\hline
\end{tabular}}}
\caption{ \label{H2energies}
       Energy and equilibrium distance of the molecule ${\rm H}_2$
       in the triplet state $^3\Pi_u$ in a magnetic field $B$ parallel to the molecular axis (this work).
       Results ${}^a$ \cite{Lai-Salpeter:1992}, ${}^b$ \cite{Lai-Salpeter:1996},
       ${}^c$ \cite{Detmer:1998} and ${}^d$ \cite{Turbiner:2010}.
       The symbol $\dagger$ indicates the results for the state $^1\Si_g$:
       $1$ \cite{et2578} and $2$ \cite{Detmer:1997}. }
\end{center}
\end{table}

\section{Conclusions}

Accurate variational calculations were carried out for the hydrogen molecule ${\rm H}_2$ with static protons for the state $^3\Pi_u$ subject to a magnetic field parallel to the molecular axis. The domain of magnetic fields studied was $5 \times 10^{10}-10^{13}\,$G.
It was shown that a 3-term compact trial function built up with physically relevant functions is already enough to provide 2-3 s.d. in energy for all magnetic fields which were studied. These results are comparable to energies obtained using larger basis sets. Note that highly accurate results for the energy of the \hh molecule are necessary to explore the stability of finite hydrogenic chains in strong magnetic fields since the \hh molecule can be present in various dissociation channels like ${\rm H}_3 \to {\rm H}_2 + {\rm H}$ or  ${\rm H}_4 \to {\rm H}_2 + {\rm H}_2$ etc. This study concludes studies of hydrogenic species needed to proceed to construct a model of hydrogenic atmosphere for neutron stars with surface magnetic field $\lesssim 10^{12}$\,G.

\section{Acknowledgements}
The research is partially supported by CONACyT grant A1-S-17364 and DGAPA grant IN113819 (Mexico).
The authors thank E. Palacios B. for technical assistance with 126-node cluster KAREN (ICN-UNAM, Mexico) where calculations were carried out.

%\bibliography{refs}

\end{document}